\documentclass[eqsecnum,showpacs,showkeys,floats,aps,nofootinbib,preprint]{revtex4}
\usepackage{bm,amsfonts, mathtools}
\usepackage{graphicx,color, wasysym}
\usepackage{textcomp}
\usepackage{amsmath,amssymb,latexsym,epsfig}

\def\de{\mathrm{d}}
\def\e{\mathrm{e}}
\def\Re{\mathrm{Re}}

\def\nn{\nonumber}

\begin{document}

\makeatletter
\title{Umbral methods and operator ordering}

\author{D. Babusci}
\email{danilo.babusci@lnf.infn.it}
\affiliation{INFN - Laboratori Nazionali di Frascati, via E. Fermi, 40, IT 00044 Frascati (Roma), Italy}

\author{G. Dattoli}
\email{dattoli@frascati.enea.it}
\affiliation{ENEA - Centro Ricerche Frascati, via E. Fermi, 45, IT 00044 Frascati (Roma), Italy}

\date{\today}


\begin{abstract}
By using methods of umbral nature, we discuss new rules concerning the operator ordering. We apply the technique of 
formal power series to take advantage from the wealth of properties of the exponential  operators. The usefulness of the obtained 
results in quantum field theory is discussed.
\end{abstract}

\maketitle

\section{Introduction}\label{Intro}
A formal power series associated with a function $f (x)$ is defined as \cite{Bell} 
\begin{equation}
\label{eq:form}
f (x) = \sum_{n = 0} \frac{c_n}{n!}\,x^n\,.
\end{equation}
If we introduce the following umbral notation \cite{Rota}
\begin{equation}
\label{eq:umb}
\hat{c}^{\,n} = c_n\,,
\end{equation}
with the assumption\footnote{Here we limit ourselves to nonnegative integer exponents (indices).} ($\hat{c}^{\,0} = 1$) 
\begin{equation}
\label{eq:cass}
\hat{c}^{\,m}\,\hat{c}^{\,n} = \hat{c}^{\,m + n} = c_{m + n}\,, 
\end{equation}
we can write the formal series \eqref{eq:form} as a pseudo-exponential
\begin{equation}
\label{eq:fx}
f (x) = \e^{\,\hat{c}\,x}\,.
\end{equation}
By taking the further freedom of defining the operator function as follows
\begin{equation}
\label{eq:Fop}
\hat{F} = \hat{F} = f (\hat{O}) = \e^{\,\hat{c}\,\hat{O}}\,,
\end{equation}
we can transform a generic operatorial function into a pseudo-exponential operator (PEO). Albeit Eq. \eqref{eq:Fop} is just a 
formal definition, we will see that the wealth of the properties of the exponential operators offers a powerful mean to transform 
the use of PEO in an effective and useful tool. If the operator $\hat{O}$ is just the ordinary derivative, we find that, as a 
consequence of Eq. \eqref{eq:umb}, the action of the exponential operator on an ordinary monomials is given by
\begin{equation}
\e^{\,\hat{c}\,\partial_x}\,x^{\,n} = (x + \hat{c})^{\,n} = \sum_{k = 0}^n \binom{n}{k}\,c_k\,x^{\,n - k}
\end{equation}
and, therefore, we can generalize the shift operator according to the following identity, valid for any function that can be 
expressed as a pseudo-exponential,
\begin{align}
g (x + \lambda\,\hat{c}) = \e^{\,\lambda\,\hat{c}\,\partial_x}\,g (x) = \e^{\,\lambda\,\hat{c}\,\partial_x}\, \e^{\,\hat{d}\,x} &= 
\sum_{n = 0}^\infty \frac{\hat{d}^{\,n}}{n!}\,\sum_{k = 0}^n \binom{n}{k}\,(\lambda\,\hat{c})^{\,k}\,x^{\,n - k}\,  \nn \\
&=  \e^{\,\hat{d}\,(x + \lambda\,\hat{c})} = \e^{\,\hat{d}\,x}\,\e^{\,\lambda\,\hat{d}\,\hat{c}}
\end{align}
where $\lambda$ is a c-number, and, in the second line, we have used the fact that the umbral operators $\hat{c}$ and 
$\hat{d}$ commute.

Along the same line of reasoning, we can consider other explicit forms of PEO. If $\hat{O} = \lambda\,x\,\partial_x$ we can 
recognize the associated exponential as a kind of dilatation operator. In this case, by setting $x = \e^t$, it's easy to show that 
\cite{Dattoli97} 
\begin{equation}\label{eq:flam}
\e^{\,\lambda\,\hat{c}\,x\,\partial_x}\,x = f (\lambda)\,x\,.
\end{equation}
In the case $\hat{O} = \lambda\,x^{\,2}\,\partial_x$, by setting $x = 1/t$, we obtain a kind of projective transformation
\begin{align}
\e^{\,\lambda\,\hat{c}\,x^{\,2}\,\partial_x}\,x &= \e^{- \lambda\,\hat{c}\,\partial_t} \left(\frac1{t}\right) = 
\frac1{t - \lambda\,\hat{c}} \nn \\
&= \frac{x}{1 - \lambda\,\hat{c}\,x}\,.
\end{align} 
i.e., if the series converges,
\begin{equation}\label{eq:qx}
\e^{\,\lambda\,\hat{c}\,x^{\,2}\,\partial_x}\,x = x\,\sum_{n = 0}^\infty c_n\,(\lambda\,x)^{\,n} = x\,q (x)\,.
\end{equation}

As an example, in the case 
\begin{equation}
\label{eq:cn}
c_n = \frac{(-1)^{\,n}}{n!}
\end{equation}
we obtain for the function $f (\lambda)$ in Eq. \eqref{eq:flam}
\begin{equation}
\label{eq:Trico}
f (\lambda) = \sum_{n = 0} \frac{(- 1)^{\,n}}{(n!)^{\,2}}\,\lambda^n\, = C_0 (\lambda) = J_0 (2\,\sqrt{\lambda}) 
\end{equation}
and for $q (x)$ in Eq. \eqref{eq:qx}
\begin{equation}
q (x) = \e^{- \lambda\,x}\,,
\end{equation}
where $C_0 (x)$ is the Bessel-Tricomi function of order zero \cite{Andrews}. The example of this function and of its power 
series expansion is paradigmatic, and, for this reason, we will study it to provide further insight into the topics we will develop in 
the following.

The formalism of PEO can be exploited along with Fourier transforms to develop new and interesting speculations. 
Given a generic function $F (x)$,  by indicating with $\tilde{F} (k)$ its Fourier transform, we can express any function of the 
type $F (\hat{c}\,x)$ as (see Eq. \eqref{eq:fx})
\begin{equation}
F (\hat{c}\,x) = \frac1{\sqrt{2\,\pi}}\,\int_{- \infty}^\infty \de k\, \tilde{F} (k)\,\e^{\,i\,k\,\hat{c}\,x} 
= \frac1{\sqrt{2\,\pi}}\,\int_{- \infty}^\infty \de k\, \tilde{F} (k)\,f (i\,k\,x)\,.
\end{equation}
that, for $c_n$ satysfing Eq. \eqref{eq:cn}, gives
\begin{equation}
F (\hat{c}\,x) = \frac1{\sqrt{2\,\pi}}\,\int_{- \infty}^\infty \de k\, \tilde{F} (k)\,C_0 (i\,k\,x)\,,
\end{equation}
This result is particularly interesting since it can be considered as a generalization of the Fourier transform itself.

In this introduction we have presented the main elements of the formalism that we are going to apply in the rest of the paper. In 
sec. \ref{UmbOpOrd} we will further develop the theory of formal series expansion, by deriving some addition theorems. Here, we will 
also study some applications of the formalism. In particular, we will discuss how the known identities of operational calculus valid for 
ordinary exponential operators can be reconsidered within the present more general framework. Finally, in sec. \ref{ConRem} we will 
analyze the obtained results in the perspective of the generalized transforms.

\section{Umbral operator ordering}\label{UmbOpOrd}
In this section we will discuss formal operation between PE's and PEO's when non commuting operators are involved. We will use 
the exponential umbral notation to state addition formulae for functions expressible as formal power series. Therefore, we consider 
(see Eq. \eqref{eq:fx}) the following identity
\begin{equation}
\label{eq:add}
f (x + y) = \e^{\,\hat{c}\,(x + y)}
\end{equation}
and ask whether an addition formula for the function $f (x)$ can be obtained.

Even though the general semi-group property $f (x + y) = f (x)\,f (y) = f (y)\,f(x)$ is lacking, the use of the identity \eqref{eq:add} allows to 
explore the possibility of a na\"{\i}ve disentanglement based on a kind of semi-group factorization. Since the operators 
$\hat{A} = \hat{c}\,x$ and $\hat{B} = \hat{c}\,y$ commute, we can write
\begin{equation}
\e^{\,\hat{c}\,(x + y)} = \e^{\,\hat{c}\,x}\,\e^{\,\hat{c}\,y} = \e^{\,\hat{c}\,y}\,\e^{\,\hat{c}\,x}
\end{equation}
that should be properly interpreted. In fact, taking into account Eq. \eqref{eq:umb}, one has
\begin{equation}
\label{eq:ecxy}
\e^{\,\hat{c}\,y}\,\e^{\,\hat{c}\,x} = \e^{\,\hat{c}\,y}\,\sum_{n = 0} \frac{c_n}{n!}\,x^n = \e^{\,\hat{c}\,y}\,f (x) 
\neq f (y + x) 
\end{equation}
The following example better clarifies the meaning of this equation. If the coefficients $c_n$ are given by Eq. \eqref{eq:cn}, 
according to Eq. \eqref{eq:cass}, one has
\begin{equation}
\e^{\,\hat{c}\,y}\,f (x) = \sum_{n = 0}^\infty \frac{(- y)^{\,n}}{n!}\,C_n (x)
\end{equation}
where 
\begin{equation}
\label{eq:Cn}
C_n (x) =  \sum_{k = 0}^\infty \frac{(- x)^{\,k}}{(n + k)!\,k!}
\end{equation}
is the Bessel-Tricomi function of order $n$. Therefore, it's clear that
\begin{equation}
\e^{\,\hat{c}\,y}\,f (x) \neq C_0 (x)\,C_0 (y)\,.
\end{equation} 
Instead, going back to Eq. \eqref{eq:ecxy} and expanding both series, we get
\begin{equation}
f (x + y) = \sum_{n = 0}^\infty \frac{\hat{c}^{\,n}\,y^{\,n}}{n!}\,\sum_{k = 0}^\infty \frac{\hat{c}^{\,k}\,x^{\,k}}{k!} = 
\sum_{n = 0}^\infty \frac{y^{\,n}}{n!}\,\varphi_n (x)
\end{equation}
where the function
\begin{equation}
\varphi_n (x) = \sum_{k = 0}^\infty \frac{c_{n + k}\,x^{\,k}}{k!} \nn
\end{equation}
in some cases can be identified with the the derivative of order $n$ of the function $f (x)$.

In the previous examples, $x$ and $y$ are $c$-numbers or commuting operators. Further problems arise if they are replaced by 
non commuting operators. We will focus our attention on the problems associated to the operator ordering, by discussing, as an example, 
the theory of SU(1,1) coherent states \cite{Louisell} from the point of view of formal power series.

By introducing two indipendent sets of creation-annihilation operators with commutation relations $(i,j = 1,2)$
\begin{equation}
[\hat{a}_i, \hat{a}_j^+] = \delta_{ij}\,, \qquad [\hat{a}_i, \hat{a}_j] = [\hat{a}_i^+, \hat{a}_j^+] = 0\,,
\end{equation}
we can define the following operators
\begin{equation}
\hat{K}_+ = \hat{a}_1^+\,\hat{a}_2^+\,,\qquad \hat{K}_- = \hat{a}_1\,\hat{a}_2\,,\qquad 
\hat{K}_0 = \frac12\,(\hat{a}_1^+\,\hat{a}_1 + \hat{a}_2\,\hat{a}_2^+)
\end{equation}
that realize a SU(1, 1) algebra with commutation brackets \cite{Barut} 
\begin{equation}
\left[\hat{K}_-, \hat{K}_+\right] = 2\,\hat{K}_0 \qquad\qquad \left[\hat{K}_0, \hat{K}_\pm\right] = \pm\,\hat{K}_\pm\,.
\end{equation}
Let us now introduce the coherent states\footnote{The name is motivated by the fact that, as we will see below, 
they are eigenstates of an annihilation operator.}
\begin{equation}
\label{eq:cohe}
| \alpha, m \rangle = \sqrt{\frac{m!}{C_m (|\alpha|^2)}}\,C_m (\alpha\,\hat{K}_+)\,| 0, m \rangle 
\qquad\qquad (\alpha \in \mathbb{C})
\end{equation}
($|0, m\rangle$ means that mode 1 is empty and mode 2 contains $m$ photons). If we define the repeated action of 
the operator $\hat{K}_+$ on the state $| 0, m \rangle$ as follows 
\begin{equation}
\hat{K}_+^{\,n}\,| 0, m \rangle = \sqrt{\frac{n!\,(n + m)!}{m!}}\,| n, n + m \rangle\,,
\end{equation}
we can write
\begin{equation}
\label{eq:amstat}
| \alpha, m \rangle = \frac1{\sqrt{C_m (|\alpha|^2)}}\,\sum_{n = 0}^\infty \frac{(- \alpha)^{\,n}}{\sqrt{n!\,(n + m)!}}\,| n, n + m\rangle\,.
\end{equation}
As a consequence of the orthonormality of the  states $| n,  m\rangle$, namely
\begin{equation}
\langle n, m | n^\prime,  m^\prime\rangle = \delta_{n, n^\prime}\,\delta_{m, m^\prime} \nn
\end{equation}
one has 
\begin{equation}
\langle \alpha, m | \alpha^\prime, m^\prime \rangle =  \frac{\delta_{m, m^\prime}}{\sqrt{C_m (|\alpha|^2)\,C_m (|\alpha^\prime|^2)}}\, 
\sum_{k = 0}^\infty \frac{(\alpha^*\,\alpha^\prime)^{\,k}}{k!\,(m + k)!}\,.
\end{equation}
Moreover, defining the action of the operator $\hat{K}_-$ as follows
\begin{equation}
\hat{K}_-\,|n, n + m\rangle = \sqrt{n\,(n + m)}\,|n - 1, n + m - 1\rangle\,,
\end{equation} 
it is easily checked that the states \eqref{eq:amstat} are eigenstates of the operator $\hat{K}_-$ 
\begin{equation}
\hat{K}_- \,| \alpha, m \rangle = - \alpha\,| \alpha, m \rangle\,.
\end{equation}

We will come back to the theory of the generalized coherent states in the last section. Here, we stress again that the reason for 
discussing these states lies in the fact that they can be defined as (we limit ourselves to the case $m = 0$) 
\begin{equation}
| \alpha, 0 \rangle = \frac1{\sqrt{C_0 (|\alpha|^2)}}\,\e^{\,\alpha\,\hat{c}\,\hat{K}_+}\,| 0, 0 \rangle 
\end{equation}
with the umbral operator $\hat{c}$ satisfying Eq. \eqref{eq:cn}. This implies that a formal series expansion can be exploited 
to treat genuine quantum problems, as those associated with two-photon interactions \cite{Walls}. 

In quantum mechanics often occur exponentials whose argument is a sum of non-commuting operators. For example, let us 
consider a formal exponential operator of the type 
\begin{equation}
\hat{E} = \exp\left\{\hat{c}\,(\Omega^*\,\hat{a}^+ + \Omega\,\hat{a})\right\}
\end{equation}
where $\Omega \in \mathbb{C}$ and $\hat{a}, \hat{a}^+$ are ladder operators, i.e. $[\hat{a}, \hat{a}^+] = 1$. The operators 
$\Omega^*\,\hat{c}\,\hat{a}^+$ and $\Omega\,\hat{c}\,\hat{a}$ satisfy the conditions for the validity of the Weyl disentanglement rule, 
according to which
\begin{equation}
\label{eq:Weyl}
\e^{\,\hat{A} + \hat{B}} = \exp\left\{- \frac{[\hat{A}, \hat{B}]}2\right\}\,\e^{\,\hat{A}}\,\e^{\,\hat{B}} 
\end{equation}
provided that $\left[\hat{A}, [\hat{B}, \hat{C}]\right] = \left[\hat{B}, [\hat{B}, \hat{C}]\right] = 0$. Therefore, we get 
\begin{equation}
\hat{E} = \exp\left\{\frac{(|\Omega|\,\hat{c})^2}2\right\}\,\e^{\Omega^*\,\hat{c}\,\hat{a}^+}\,\e^{\Omega\,\hat{c}\,\hat{a}}\,.
\end{equation}
Let us now consider the action of the above operator on the vacuum state of the electromagnetic field, namely\footnote{Notice 
that for $\hat{c} = 1$ the states $| \Omega \rangle$ reduces to the ordinary Glauber states.} 
\begin{equation}
\label{eq:Ome}
| \Omega \rangle = \hat{E}\,| 0 \rangle =\exp\left\{\frac{(|\Omega|\,\hat{c})^2}2\right\}\,\e^{\,\Omega^*\,\hat{c}\,\hat{a}^+}
\,| 0 \rangle\,. 
\end{equation}
It is easily checked that
\begin{equation}
\hat{a}\,| \Omega \rangle = \Omega^*\,\hat{c}\,| \Omega \rangle\,,
\end{equation}
i.e., that the states \eqref{eq:Ome} are not, in general, coherent. The use of the generating function of two-variable Hermite 
polynomials
\begin{equation}
\sum_{n = 0}^\infty \frac{t^{\,n}}{n!}\,H_n (x, y) = \e^{\,x\,t + y\,t^2} \qquad\qquad H_n (x, y) = n! \sum_{k = 0}^{[n/2]} 
\frac{x^{\,n - 2 k}\, y^{\,k}}{(n - 2 k)!\,k!}
\end{equation}
allows us to write  
\begin{equation}
| \Omega \rangle = \sum_{n = 0}^\infty \frac{\hat{c}^{\,n}}{n!}\,| h_n \rangle
\end{equation}
where, since $(a^+)^k | 0 \rangle = \sqrt{k!}\,| k \rangle$
\begin{equation}
| h_n \rangle = H_n \left(\Omega^*\,\hat{c}\,\hat{a}^+, \frac{|\Omega|^2}2\right)\,| 0 \rangle = n! \sum_{k = 0}^{[n/2]} 
\frac{(\Omega^*)^{n - 2 k}\,|\Omega|^{2 k}}{2^k\,k!\,\sqrt{(n - 2 k)!}}\,| n - 2 k \rangle\,.
\end{equation}
These states are called \textit{Hermite quantum states}. They are not mutually orthogonal
\begin{equation}
\langle h_n | h_m \rangle = \frac{n!\,m!}{\sqrt{2^{\,n - m}}}\,| \Omega |^{\,n + m}\,\sum_{k = 0}^{[n/2]} 
\frac1{4^k\,k!\,\left(\displaystyle \frac{n - m}2 + k\right)!\,(n - 2 k)!}\,,
\end{equation}
and the scalar product with the states $| m \rangle$ is given by
\begin{equation}
\langle m | h_n \rangle = \frac{n!}{\sqrt{2^{\,n - m}}}\,\frac{(\Omega^*)^{m}\,| \Omega |^{\,n - m}}
{\displaystyle \left(\frac{n - m}2\right)!\,\sqrt{m!}}\,.
\end{equation}

\section{Further examples and conclusions}\label{ConRem}
The methods we have proposed in this paper are useful to deal with operator functions, but can also be exploited  
to treat many other problems, and this concluding section will be devoted to discuss some examples of them.

As a first example, we will consider the following evolution equation
\begin{equation}
\label{eq:evol}
\hat{\Delta}\,\Psi (x, \delta) = \hat{H}\,\Psi (x, \delta)
\end{equation}
where the eigenvalue equation for the operator $\hat{\Delta}$ is
\begin{equation}
\hat{\Delta}\,E (\lambda\,\delta)  = \lambda\,E (\lambda\,\delta)
\end{equation}
with $E (\alpha) = \e^{\,\alpha\,\hat{c}}$. The evolution operator associated to Eq. \eqref{eq:evol} is
\begin{equation}
\hat{U} (\delta) = \hat{E} (\delta\,\hat{H}) = \e^{\,\delta\,\hat{c}\,\hat{H}}\,.
\end{equation}
In the case $\hat{H} = \partial_x^2$ and $\Psi (x, 0) = \e^{- x^2}$, the solution of the problem \eqref{eq:evol} can be written as 
a formal restatement of the Glaisher identity \cite{Book} as follows
\begin{align}
\label{eq:glais}
\Psi (x, \delta) = \e^{\,\delta\,\hat{c}\,\partial_x^2}\,\e^{- x^2} &= \frac1{\sqrt{1 + 4\,\delta\,\hat{c}}}\,
\exp\left\{- \frac{x^2}{1 + 4\,\delta\,\hat{c}}\right\} \nn \\
&= \sum_{n = 0}^\infty (4\,\delta\,\hat{c})^n\,L_n ^{- \frac12} (x^2)
\end{align}
where in the second line an expansion involving the Laguerre polynomials $L_n$ has been used. The relevance of the previous 
results, along with the formalism of the Laguerre derivative \cite{Book}, will be discussed later in this section.

The method proposed can also usefully applied to solve partial differential equations. We consider indeed the following partial 
differential equation
\begin{equation}
\label{eq:Lagu}
\partial_t\,t\,\partial_t\,F (x, t) = - \partial_x^2\,F (x, t) \qquad\qquad F (x, 0) = g (x)
\end{equation}
that we call pseudo-heat equation. Since the Bessel-Tricomi function of order zero is an eigenfunction of the Laguerre 
derivative operator $_L\hat{D}_\xi = - \partial_\xi\,\xi\,\partial_\xi$ \cite{Book}, i.e. 
\begin{equation}
_L\hat{D}_\xi\,C_0 (\xi) = \lambda\,C_0 (\xi)\,, \nn
\end{equation}
according to Eq. \eqref{eq:Trico} we can write the solution of Eq. \eqref{eq:Lagu} in the form
\begin{equation}
F (x, t) = C_0 (t\,\partial_x^2)\,g (x) = \e^{-\hat{c}\,t\,\partial_x^2}\,g (x)\,.
\end{equation}
The ÒrulesÓ for use in applications of the evolution operator $\hat{U} (t) = \e^{-\hat{c}\,t\,\partial_x^2}$ associated to Eq. \eqref{eq:Lagu} 
deserve an accurate treatment, going beyond the scope of this paper. Here we only note that in the case $g (x) = x^m$, the solution 
of the Eq. \eqref{eq:Lagu} is given by
\begin{equation}
F (x, t) = \sum_{n = 0}^\infty \frac{(- t\,\hat{c})^{\,n}}{n!}\,\partial_x^{\,2\,n}\,x^m = \,_LH_m (x, - t)
\end{equation}
where
\begin{equation}
_L H_m (\xi, \tau) = m!\,\sum_{k = 0}^{[m/2]} \frac{\tau^{\,m}\,\xi^{\,m - 2 k}}{(k!)^2\,(m - 2 k)!}
\end{equation}
are sometimes called \textit{hybrid} polynomials because their properties are intermediate between those of Laguerre and Hermite 
polynomials \cite{Book}. They can be framed within the context of the formalism of the hypergeometric function. However their properties 
are so peculiar, interesting and useful in applications that they should be considered separately.

In closing the paper we discuss the combination of the PEO formalism with integral transforms method. Let us consider the following identity 
$(\Re\,\nu > 0)$
\begin{equation}
\label{eq:Lapl}
\frac1{(1 + \hat{c}\,x)^{\,\nu}} = \frac1{\Gamma (\nu)}\,\int_0^\infty \de s\,\e^{- s\,(1 + \hat{c}\,x)}\,s^{\,\nu - 1}\,.
\end{equation}
The use of Eq. \eqref{eq:fx} allows us to recast it in the form
\begin{equation}
\frac1{(1 + \hat{c}\,x)^{\,\nu}} = \frac1{\Gamma (\nu)}\,\int_0^\infty \de s\,\e^{- s}\, f(- s\,x)\,s^{\,\nu - 1}\,,
\end{equation}
that, in some sense, generalizes the Laplace and Mellin transforms. For example, in the case
\begin{equation}
\hat{c}^{\,n} = \frac1{(m\,n + p)!}  
\end{equation}
from Eq. \eqref{eq:Lapl} one has
\begin{equation}
\frac1{(1 + \hat{c}\,x)^{\,\nu}} = \frac1{\Gamma (\nu)}\,\sum_{k = 0}^\infty \frac{(- 1)^{\,k}\,x^{\,k}}{k!\,(m\,k + p)!}\,\Gamma (k + \nu)\,.
\end{equation}

The theory of PEO Laplace and Fourier transform deserves a deeper treatment to which will be devoted a forthcoming paper.

\end{document}